\documentclass[preprint,11pt]{elsarticle}
\usepackage{color,amssymb,subfigure,graphicx,amsmath,slashbox,multirow,multicol}
\usepackage{psfrag}
\usepackage{algorithm}
\usepackage{algorithmic}
\journal{arXiv}
\begin{document}
\graphicspath{{figures/}}
\newcommand{\sps}{\scriptsize}
\newcommand{\eqs}{\normalsize}
\begin{frontmatter}
\title{Deep Neural Network for Analysis of DNA Methylation Data}
  \author{ Hong Yu and Zhanyu Ma}
 \address{Pattern Recognition and Intelligent System Laboratory\\
  Beijing University of Posts and Telecommunications, Beijing, China.}
\begin{abstract}
Many researches demonstrated that the DNA methylation, which occurs in the context of a CpG, has strong correlation with diseases, including cancer. There is a strong interest in analyzing the DNA methylation data to find how to distinguish different subtypes of the tumor. However, the conventional statistical methods are not suitable for analyzing the highly dimensional DNA methylation data with bounded support. In order to explicitly capture the properties of the data, we design a deep neural network, which composes of several stacked binary restricted Boltzmann machines, to learn the low dimensional deep features of the DNA methylation data. Experiments show these features perform best in breast cancer DNA methylation data cluster analysis, comparing with some state-of-the-art methods.
\end{abstract}

\end{frontmatter}

\section{Introduction}

DNA methylation, occurring in the context of a CpG dinucleotide, is a kind of epigenetic modification in human genome, which can be inherited through cell division~\cite{S1}. Some special methylation patterns are found in many genetic diseases including various types of cancer~\cite{S2}. Due to the role of methylation patterns in the etiology of complex diseases, DNA methylation analysis becomes a powerful tool in cancer diagnosis, treatment, and prognostication.

The high throughput methylation profiling technology,~\emph{e.g.}, the Illumina methylation platform, has been developed to survey methylation status of more than $1500$ CpG sites associated with over $800$ cancer-related genes\cite{S3}, which makes it is easily to measure genome-wide from limited amounts of DNA and allows measurements in clinical specimens~\cite{S4}.
Currently, there is a strong interest in studying how the methylation profiles can be used to distinguish different subtypes of the tumor. These researches perform unsupervised clustering of large-scale DNA methylation data sets~\cite{S5}.

Formerly, the clustering work focused on the sequence level. Recently, the exact value levels of methylation expression has been fully considered and attracts more and more attentions. The beta-value, which means the ratio of the methylated probe intensity and the overall intensity (sum of methylated and unmethylated probe intensities), is usually used to express the methylation level~\cite{S6}. Beta-value is quantified naturally bounded between $0$ and $1$. Under ideal conditions, a value of $0$ means the CpG site is completely unmethylated and the value of $1$ indicates the site is fully methylated~\cite{S6}.

Due to the unique non-Gaussian characteristics, traditional Gaussian distribution-based clustering methods are not be appropriate for DNA methylation data analysis~\cite{S7}. In order to make the methylation data range satisfy the definition of Gaussian distribution, some transformation methods were proposed, such as the M-value method which uses the logit-transform to change the feature range to $(0,\infty)$~\cite{S6}. However, these transformations may cause worse inference and the mathematical transformation on methylation data is lack of the support of the biological significance~\cite{S7}. Hence, more researchers tend to deal with the beta-value directly by using beta mixture model~\cite{S6,S33,S8,S9}. The DNA methylation expressions can be modeled by a mixture of beta distributions. Unfortunately, estimation of the beta mixture model does not have an analytically tractable solution and the analysis based on it cannot be derived in an explicit form. Thus, many approximate solution methods, such as variational Bayes inference~\cite{S10} and Gibbs sampling~\cite{S11}, have been adopted to solve this problem.

Moreover, the extremely high dimensions of the methylation data also yield many practical problems in pattern analysis. Generally speaking, we conduct dimensionality reduction work before cluster analysis. However, the traditional method,~\emph{e.g.}, principal component analysis (PCA)~\cite{S13} and nonnegative matrix factorization (NMF)~\cite{S14}, are all based on the Gaussian distribution assumption. The bounded support property cannot be preserved during the transformation. Hence, the statistical properties of the non-Gaussian DNA methylation data sets cannot be efficiently described by these existing dimension reduction methods.

Recently, deep learning algorithms, which are making artificial intelligence smarter for vision and speech, have been used to solve the bioinformatic problems. It provides an efficient tool for analyzing considerable high dimensional biomedical data. By unsupervised or semi-supervised training, an interactional multilayer complex neural networks can be build to automatically extract the unobserved deep information hidden in the mussy biomedical data. The deep neural network has been applied to predict how genes were spliced together in mice~\cite{b1}, assess the state of Parkinson＊s Disease~\cite{b2}, denoise the ECG signals, and many other biomedical researches~\cite{b3}.

Inspired by these work, we use an auto-encode deep neural network model to carry out the dimensionality reduction task in this paper~\cite{S15}. The network is composed by several stacked binary restricted Boltzmann machines (RBM). The RBM is a probabilistic graphical model that can be interpreted as stochastic neural network~\cite{S16}, the input and output of binary RBM are regarded as probabilistic values and naturally bounded in $[0,1]$. Such properties have similar manner with the DNA methylation data. Thus, this model is suitable to analyze the methylation data.

In the experimental part, we examine the effect of DNA methylation data analysis based on deep neural network (DNN). Firstly, we check the dimension reduction effect. By high-dimensional data visualization technology, we can observe that the features with low dimension can distinguish cancer from normal samples efficiently. Secondly, we conduct some unsupervised clustering analysis using the features extracted from the DNN model. The results demonstrate that cancer and healthy samples can be efficiently clustered into different groups based on the DNN-based features.

The rest of the paper is organized as follow: we introduce the adopted DNN structure in Section $2$. The experimental results are presented in section $3$, in which we describe the DNA methylation data sets and illustrate the DNA data analysis results. Finally, we draw some conclusions in section $4$.


\section{Deep Neural Networks Structure for DNA Methylation Data Analysis}
\label{System}
Since $2006$, deep learning method has been widely used in many fields of science researches~\cite{S17,S18,S19}. Many different deep learning models are applied for different applications,~\emph{e.g.}, deep neural networks (DNN) and convolutional neural network (CNN)~\cite{S20}. DNN has a classical auto encode (AE) structure~\cite{S16} which can be used to extract deep feature automatically. An AE is composed with several stacked RBM, as shown in ${\bf{Fig.1}}$. (a).
\begin{figure}[!t]
 \centering
  \subfigure[\label{Subfig: Strong} Deep neural networks structure.]{\includegraphics[width=.45\textwidth]{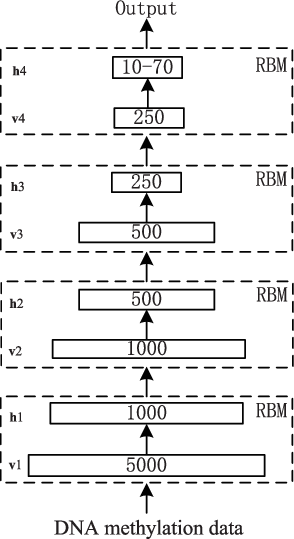}}\hspace{4mm}
  \subfigure[\label{Subfig: Strong} Inner structure of RBM.]{\includegraphics[width=.45\textwidth]{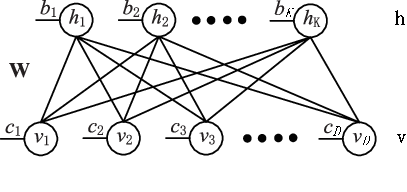}}\hspace{2mm}
 \caption{Illustration of a stacked RBM.  }
 \label{fig:fig1}
\end{figure}
The RBM is an undirected graphical model that defines the distribution of visible units using binary hidden units. The inner structure of RBM is shown in  ${\bf{Fig.1}}$. (b). The joint distribution of binary visible units and binary hidden units is written as follows:
\begin{equation}\label{eq1}%
{\mathop{\rm P}\nolimits} ({\bf{v}},{\bf{h}}) = \frac{1}{Z}\exp ( - {\mathop{\rm E}\nolimits} ({\bf{v}},{\bf{h}})),
\end{equation}%
\begin{equation}\label{eq2}%
{\mathop{\rm E}\nolimits} ({\bf{v}},{\bf{h}}) =  - \sum\limits_{i = 1}^D {\sum\limits_{k = 1}^K {{v_i}{W_{ik}}{h_k}}  - \sum\limits_{k = 1}^K {{b_k}{h_k}}  - } \sum\limits_{i = 1}^D {{c_i}{v_i}},
\end{equation}
where ${\bf{v}} \in {{\rm{\{ 0,1\} }}^D}$ are the visible (input) units, and ${\bf{h}} \in {{\rm{\{ 0,1\} }}^K}$ are the hidden (output) units. Z is the normalizing constant, and ${\bf{W}} \in {R^{D \times K}}$, ${\bf{b}} \in {R^K}$, ${\bf{c}} \in {R^D}$ are the weight matrix, hidden, and visible bias vectors, respectively~\cite{S21}.

\begin{table*}[!t]
\caption{Comparisons the clustering effect based on different feature extraction and unsupervised clustering methods.}
\centering
\begin{tabular}{|c|c|c|c|}

 \hline
Method &  Error rate(\%) & Cancer$\rightarrow$Healthy & Healthy $\rightarrow$Cancer  \\
\hline
\hline
PCA+ k-means &5.15 & 7 & 0 \\
\hline
PCA+ GMM &5.88 & 8 & 0 \\
\hline
PCA+ SOM & 4.41 & 0 & 6 \\
\hline
\hline
NMF + k-means & 8.82 & 12 & 0 \\
\hline
NMF + GMM & 12.5 & 17 & 0 \\
\hline
NMF + SOM & 3.68 & 1 & 4 \\
\hline
\hline
DNN + k-means & 5.15 & 7 & 0 \\
\hline
DNN + GMM & 5.15 & 7 & 0 \\
\hline
\emph{\bf{DNN+SOM}} & $\bf{2.94}$ & $\bf{4}$ & $\bf{0}$  \\
\hline
\end{tabular}
\end{table*}
Since there are no connections between the units in the same layer, visible units are mutually conditionally independent given the hidden units, and vice versa~\cite{S21}. The conditional probabilities of the RBM can be written as follows:
\begin{equation}\label{eq3}%
P({v_i} = 1|h) = \sigma (\sum\limits_k {{W_{ik}}{h_k} + {c_i}} ),
\end{equation}%
\begin{equation}\label{eq4}%
P({h_k} = 1|v) = \sigma (\sum\limits_k {{W_{ik}}{v_i} + {b_k}} ),
\end{equation}%
where
\begin{equation}\label{eq5}%
\sigma (x) = \frac{1}{{1 + {e^{ - x}}}}.
\end{equation}%
Training the RBM corresponds to maximizing the log-likelihood of the data with respect to parameters$\left\{ {{\bf{W}}{\rm{,}}{\bf{b}}{\rm{,}}{\bf{c}}} \right\}$. Although the gradient is intractable to compute, contrastive divergence~\cite{S16} can be used to approximate it.

In training process, each RBM can be trained independently and the output of the bottom RBM can be used as the input data of the upper RBM. After getting all the parameters for each layer, we can get a DNN as shown in $\bf{Fig. 1}$. (a) to carry out feature extraction and dimension reduction.

Because the input unit is binary, the input data sent to the input unit should be bounded from $0$ to $1$. This peculiarity just adapts the characteristic of DNA methylation data. The output getting from equation~\ref{eq4} are also bounded, which makes the features finally extracted from the top layer are still in accordance with the characteristics of methylation data (in $[0,1]$).

In this paper we built a DNN with $4$ layers. The (input, output) unit numbers of the bottom $3$ layers were set as $\{(5000, 1000), (1000, 500), (500, 250)\}$, respectively. The top layer's output units number was set as $\{10, 20, 30, 40, 50,$ $60, 70\}$ to extract the best feature of the input data. The analysis of results are shown in Section $3$.

\section{Experimental Results and Discussions}
\label{System}

The DNA methylation data were obtained from the Gene Expression Omnibus (GEO) website~\cite{S22}. GEO is a public functional genomics data repository supporting MIAME-compliant data submissions. In this section, we used the dataset GSE$32393$ to evaluate the dimension reduction and unsupervised clustering effect~\cite{S23}. This dataset include $136$ women breast tissues samples with $113$ breast cancers samples and $23$ non-neoplastic samples. From each sample, we obtained approximately $27,578$ DNA methylation features. In order to reduce the interference of the noise data we selected $5,000$ features with the highest variance across the $136$ samples as the experimental data.

The selected $5,000$ dimensional features were sent to the DNN described in Section $2$ to realize further dimension reduction. In order to find the best feature dimension, we set the output units number as $\left\{10, 20, 30, 40, 50, \protect\\ 60, 70\right\}$. Then these dimension reduced features were clustered by the self-organizing feature maps (SOM) methods~\cite{S25}. After intensive experiments, we found out when setting the cluster number $k=5$ and the feature dimension reduced to $30$, the best results can be achieved.

Comparing the dimension reduced effect between DNN and some other traditional methods, we also reduce the $5,000$ features into $30$ dimension using traditional PCA and NMF method~\cite{S26}, respectively.

In order to visually assess the effect of feature selection,  we adopted the t-distributed stochastic neighbor embedding (t-SNE) method to visualize the high-dimensional data in two-dimensional space~\cite{S24}. t-SNE is a nonlinear dimensionality reduction technique which has been used in a wide range of applications, including computer security research,~\cite{T1} music analysis~\cite{T2}, cancer research~\cite{T3}, and bioinformatics~\cite{T4}. It is particularly well suited for embedding high-dimensional data into a space of two or three dimensions, which can then be visualized by a scatter plot. Especially, it models each high-dimensional object by a two- or three-dimensional point in such a way that similar objects are modeled by nearby points and dissimilar objects are modeled by distant points.
\begin{figure*}[!t]
 \centering
          \subfigure[\label{Subfig: Weak}]{\includegraphics[width=.45\textwidth]{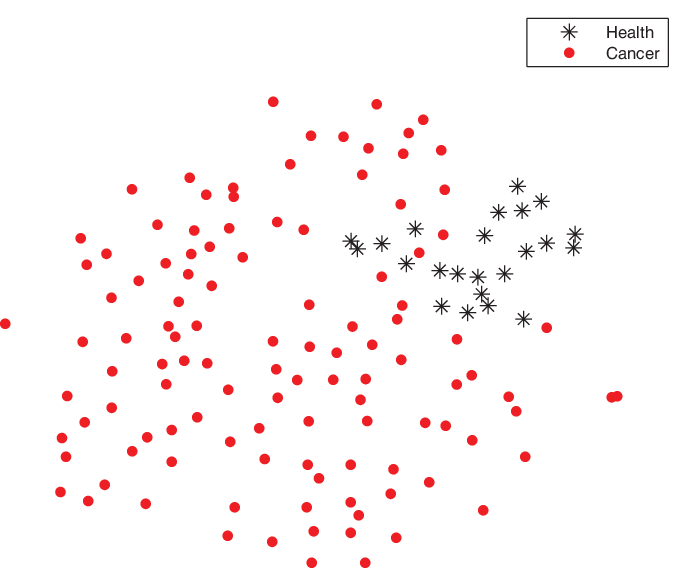}}\hspace{1mm}
          \subfigure[\label{Subfig: Strong}]{\includegraphics[width=.45\textwidth]{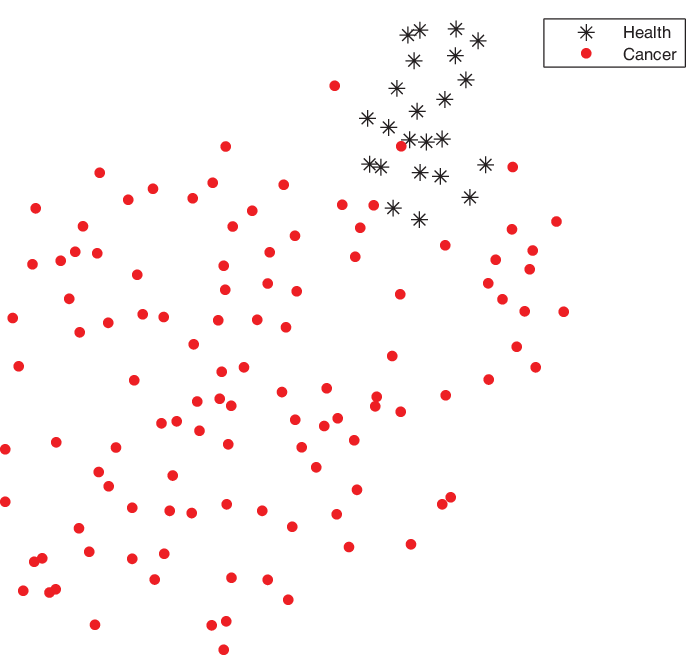}}\hspace{1mm}
                    \subfigure[\label{Subfig: Weak}]{\includegraphics[width=.45\textwidth]{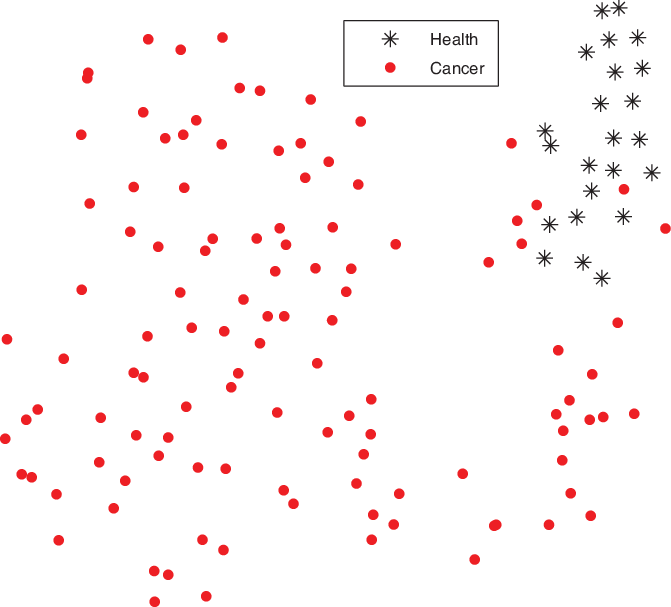}}
                    \vspace{2mm}
          \subfigure[\label{Subfig: Strong}]{\includegraphics[width=.45\textwidth]{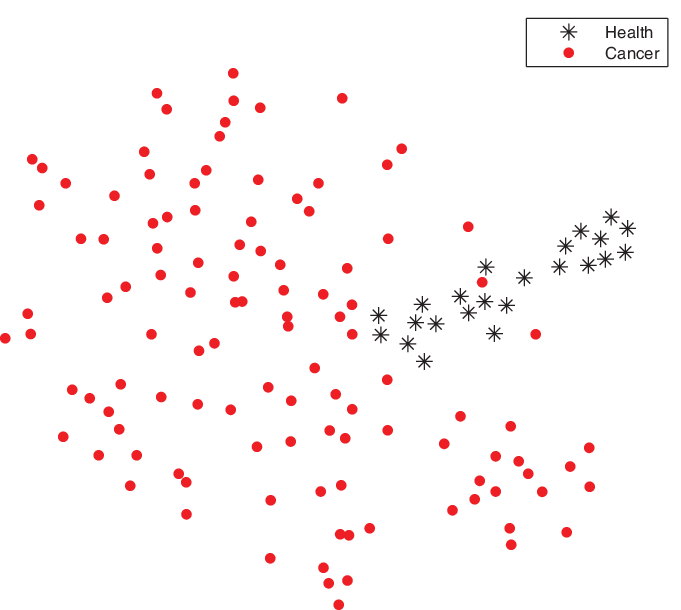}}\hspace{1mm}
                    \subfigure[\label{Subfig: Weak}]{\includegraphics[width=.45\textwidth]{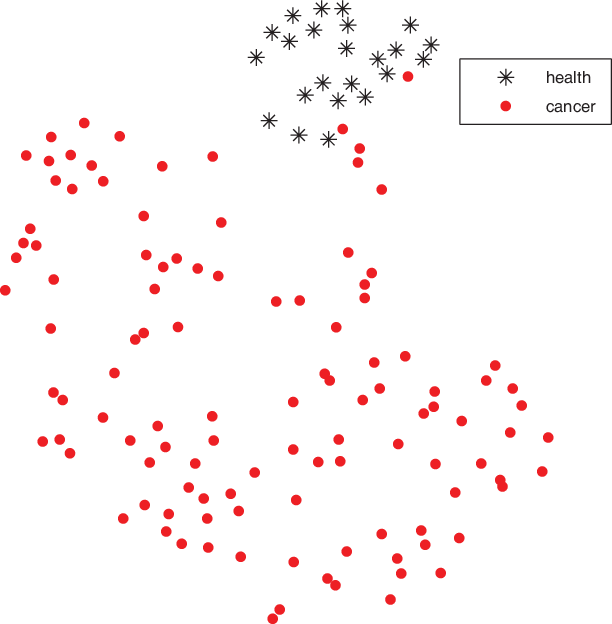}}\hspace{1mm}
          \subfigure[\label{Subfig: Strong}]{\includegraphics[width=.45\textwidth]{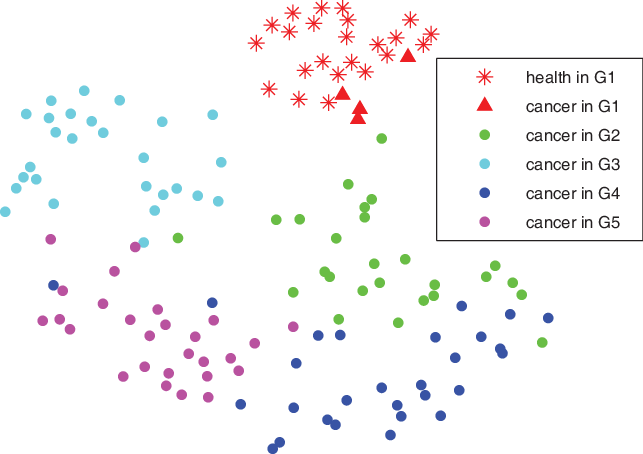}}
 \caption{(a) Visualization of the original $27,578$ dimensional data; (b) Visualization of the selected $5,000$ dimensional data; (c) The visualization of $30$ dimensional features extracted by PCA method; (d) The visualization of $30$ dimensional features extracted by NMF method~\cite{S26}; (e) The visualization of $30$ dimensional features extracted by DNN; (f) Clustering result of the $30$ dimensional DNN features based on SOM method.
  }
 \label{fig:fig1}
\end{figure*}


\begin{table*}[!t]
\centering
\caption{Comparisons the clustering results between DNN+SOM method and some other mixture model-based methods.}

\small
\begin{tabular}{|c|c|c|c|}
\hline
Method &  Error Rate(\%) & Cancer$\rightarrow$Healthy & Healthy $\rightarrow$Cancer  \\
\hline
\hline
PCA + VBGMM &6.62 & 9 & 0 \\
\hline
BGNMF+ RPBMM &3.68 & 4 & 1 \\
\hline
BGNMF+ VBBMM & 3.68 & 4 & 1 \\
\hline
SC + VBWMM & 5.15 & 7 & 0 \\
\hline
\bf{DNN+SOM} & $\bf{2.94}$ & $\bf{4}$ & $\bf{0}$  \\
\hline
\end{tabular}
\end{table*}
The visualization results are shown in $\bf{Fig. 2}$ and it is clearly illustrated that the cancer (red circle) and healthy (black star) points in $\bf{Fig. 2}$. (b) are better separated than those in $\bf{Fig. 2}$. (a). This indicates that feature selection can potentially improve the clustering performance. $\bf{Fig. 2}$. (c), (d) and (e) showed the dimensional reduced effect of PCA, NMF, and DNN, respectively. It was shown clear that the DNN method can capture the features of DNA methylation data better than the PCA and NMF method, which means that the DNN features are more effective for the clustering analysis.

In order to make fair comparisons, we also applied the k-means, the Gaussian mixture model (GMM), and the SOM method to cluster the features extracted by PCA, NMF, and DNN methods, respectively. The results were shown in $\bf{Table}$ 1. It could be observed that the DNN features performs the best among all the clustering methods, and the SOM clustering method performed better than the others.

$\bf{Figure~2}$. (f) shows the visualization clustering result of the $30$ dimensional DNN features based on the SOM method. All the samples were clustered into $5$ groups and all the $23$ healthy samples (red star) were grouped in group $1$ (the healthy group, denoted in red), only $4$ cancer samples (red triangle) were allocated in this group by mistake. The group $2, 3, 4$ and $5$ were cancer group and none of the healthy samples was clustered in them.

We also compared the DNN+SOM method with some other mixture model-based methods including PCA+VBGMM (variational Bayesian Gaussian mixture model)~\cite{S10}, BG-NMF (Beta-Gamma Nonnegative matrix factorization) + RPBMM (recursive partitioning Beta mixture model)~\cite{S26}, BG-NMF + VBBMM (variational Bayesian estimation framework for Beta mixture model)~\cite{S8} and SC (spectral clustering) + VBWMM (variational Bayesian estimation of Watson mixture model)~\cite{S7}. The result was shown in $\bf{Table~2}$ (some result were from paper~\cite{S7}). We can see that, compared with the complex probabilistic mixture models, the DNN is a simple yet effective model which can get the best result among all the compared methods.

\section{Conclusions}
\label{System}
The DNA methylation level can be used to distinguish cancer gene from normal gene. However, the high dimensionality makes it hard to be analyzed directly and analyzed. Meanwhile, the non-Gaussian properties of the data make many conventional dimension reduction methods do not work well. In this paper, we adopted a dimension reduction method based on deep neural networks (DNN). The DNN is built by $4$ stacked binary restricted Boltzmann machines (RBM). The binary input and output units of the RBM fits the DNA methylation data's bounded support property well and the self-learning ability can extract the low dimensional features automatically. The experiment results demonstrate the low dimensional features getting from DNN can separate the normal samples from the cancer samples effectively. Compared with some recently proposed probabilistic mixture model-based methods, the DNN-based method shows significant advantages.


\begin{thebibliography}{1}
\bibitem{S1}
A.~M.~Deaton and A.~Bird, ``CpG islands and the regulation of transcription'',~\emph{Genes Dev.} No. 25, pp. 1010-1022, 2011.

\bibitem{01}
B.~Everitt and D.~Hand, \emph{Finite Mixture Distributions}.\hskip 1em plus
  0.5em minus 0.4em\relax Chapman and Hall, London, UK, 1981.

\bibitem{02}
G.~McLachlan and D.~Peel, \emph{Finite Mixture Models}.\hskip 1em plus 0.5em
  minus 0.4em\relax New York, NY, USA: Wiley, 2000.

\bibitem{28}
C.~M. Bishop, \emph{Pattern Recognition and Machine Learning (Information
  Science and Statistics)}.\hskip 1em plus 0.5em minus 0.4em\relax
  Springer-Verlag New York, Inc., 2006.

\bibitem{03}
N.~Bouguila, D.~Ziou, and J.~Vaillancourt, ``Unsupervised learning of a finite
  mixture model based on the {D}irichlet distribution and its application,''
  \emph{IEEE Transactions on Image Processing}, vol.~13, no.~11, pp.
  1533--1543, 2004.

\bibitem{Reynolds1995}
D.~A. Reynolds and R.~C. Rose, ``Robust text-independent speaker identification
  using {G}aussian mixture speaker models,'' \emph{IEEE Transactions on Speech
  and Audio Processing}, vol.~3, no.~1, pp. 72--83, 1995.

\bibitem{Nasios2006}
N.~Nasios and A.~G. Bors, ``Variational learning for {G}aussian mixture
  models,'' \emph{IEEE Transactions on Systems Man and Cybernetics Part B
  (Cybernetics)}, vol.~36, no.~4, pp. 849--862, July 2006.

\bibitem{Jung2014}
J.~Jung, S.~R. Lee, H.~Park, S.~Lee, and I.~Lee, ``Capacity and error
  probability analysis of diversity reception schemes over generalized-${K}$
  fading channels using a mixture gamma distribution,'' \emph{IEEE Transactions
  on Wireless Communications}, vol.~13, no.~9, pp. 4721--4730, Sept 2014.

\bibitem{Houseman2008}
E.~A. Houseman, B.~C. Christensen, R.~F. Yeh, C.~J. Marsit, M.~R. Karagas,
  M.~Wrensch, H.~H. Nelson, J.~Wiemels, S.~Zheng, J.~K. Wiencke, and K.~T.
  Kelsey, ``Model-based clustering of {DNA} methylation array data: a
  recursive-partitioning algorithm for high-dimensional data arising as a
  mixture of beta distributions,'' \emph{Bioinformatics}, vol.~9, p. 365, 2008.

\bibitem{Nascimento2012}
J.~M.~P. Nascimento and J.~M. Bioucas-Dias, ``Hyperspectral unmixing based on
  mixtures of {D}irichlet components,'' \emph{IEEE Transactions on Geoscience
  and Remote Sensing}, vol.~50, no.~3, pp. 863--878, March 2012.



\bibitem{He2010}
Q.~He, K.~Chang, E.~P. Lim, and A.~Banerjee, ``Keep it simple with time: A
  reexamination of probabilistic topic detection models,'' \emph{IEEE
  Transactions on Pattern Analysis and Machine Intelligence}, vol.~32, no.~10,
  pp. 1795--1808, Oct 2010.

\bibitem{10}
T.~Bdiri and N.~Bouguila, ``Positive vectors clustering using inverted
  {D}irichlet finite mixture models,'' \emph{Expert Systems with Applications},
  vol.~39, no.~2, pp. 1869--1882, 2012.

\bibitem{Bdiri2013}
------, ``Bayesian learning of inverted {D}irichlet mixtures for {SVM} kernels
  generation,'' \emph{Neural Computing and Applications}, vol.~23, no.~5, pp.
  1443--1458, 2013.

\bibitem{Bdiri2013a}
T.~Bdiri, N.~Bouguila, and D.~Ziou, ``Visual scenes categorization using a
  flexible hierarchical mixture model supporting users ontology,'' in
  \emph{IEEE International Conference on TOOLS with Artificial Intelligence},
  2013, pp. 262--267.

\bibitem{Markley2010}
S.~C. Markley and D.~J. Miller, ``Joint parsimonious modeling and model order
  selection for multivariate {G}aussian mixtures,'' \emph{IEEE Journal of
  Selected Topics in Signal Processing}, vol.~4, no.~3, pp. 548--559, June
  2010.

\bibitem{11}
Z.~Liang and S.~Wang, ``An {EM} approach to {MAP} solution of segmenting tissue
  mixtures: a numerical analysis.'' \emph{IEEE Transactions on Medical
  Imaging}, vol.~28, no.~2, pp. 297--310, 2009.

\bibitem{12}
N.~Bouguila and D.~Ziou, ``Unsupervised selection of a finite {D}irichlet
  mixture model: an {MML}-based approach,'' \emph{IEEE Transactions on
  Knowledge and Data Engineering}, vol.~18, no.~8, pp. 993--1009, June 2006.

\bibitem{Richardson1996}
S.~Richardson and P.~J. Green, ``Corrigendum: On bayesian analysis of mixtures
  with an unknown number of components,'' \emph{Journal of the Royal
  Statistical Society}, vol.~60, no.~3, p. 661, 1996.

\bibitem{Huang2016}
L.~Huang, Y.~Xiao, K.~Liu, H.~C. So, and J.~K. Zhang, ``Bayesian information
  criterion for source enumeration in large-scale adaptive antenna array,''
  \emph{IEEE Transactions on Vehicular Technology}, vol.~65, no.~5, pp.
  3018--3032, May 2016.

\bibitem{Chen2013}
X.~Chen, ``Using {A}kaike information criterion for selecting the field
  distribution in a reverberation chamber,'' \emph{IEEE Transactions on
  Electromagnetic Compatibility}, vol.~55, no.~4, pp. 664--670, Aug 2013.

\bibitem{Bousmalis2015}
K.~Bousmalis, S.~Zafeiriou, L.~P. Morency, M.~Pantic, and Z.~Ghahramani,
  ``Variational infinite hidden conditional random fields,'' \emph{IEEE
  Transactions on Pattern Analysis and Machine Intelligence}, vol.~37, no.~9,
  pp. 1917--1929, Sept 2015.

\bibitem{Meila2016}
M.~Meil\v{a} and H.~Chen, ``Bayesian non-parametric clustering of ranking
  data,'' \emph{IEEE Transactions on Pattern Analysis and Machine
  Intelligence}, vol.~38, no.~11, pp. 2156--2169, Nov 2016.

\bibitem{Xu2016}
Y.~Xu, M.~Megjhani, K.~Trett, W.~Shain, B.~Roysam, and Z.~Han, ``Unsupervised
  profiling of microglial arbor morphologies and distribution using a
  nonparametric {B}ayesian approach,'' \emph{IEEE Journal of Selected Topics in
  Signal Processing}, vol.~10, no.~1, pp. 115--129, Feb 2016.

\bibitem{18}
T.~S. Ferguson, ``A {B}ayesian analysis of some nonparametric problems,''
  \emph{Annals of Statistics}, vol.~1, no.~2, pp. 209--230, 1973.

\bibitem{19}
C.~E. Antoniak, ``Mixtures of {D}irichlet processes with applications to
  {B}ayesian nonparametric problems,'' \emph{Annals of Statistics}, vol.~2,
  no.~6, pp. 1152--1174, 1974.


\bibitem{Hjort2010}
N.~L. Hjort, C.~Holmes, P.~M\"{u}ller, and S.~G. Walker, Eds., \emph{Bayesian
  Nonparametrics}.\hskip 1em plus 0.5em minus 0.4em\relax Cambridge University
  Press, 2010.

\bibitem{20}
Y.~W. Teh and D.~M. Blei, ``Hierarchical {D}irichlet processes,'' \emph{Journal
  of the American Statistical Association}, vol. 101, no. 476, pp. 1566--1581,
  2006.

\bibitem{Foti2015}
N.~J. Foti and S.~A. Williamson, ``A survey of non-exchangeable priors for
  {B}ayesian nonparametric models,'' \emph{IEEE Transactions on Pattern
  Analysis and Machine Intelligence}, vol.~37, no.~2, pp. 359--371, Feb 2015.

\bibitem{17}
W.~Fan and N.~Bouguila, ``Online learning of a {D}irichlet process mixture of
  beta-{L}iouville distributions via variational inference,'' \emph{IEEE
  Transactions on Neural Networks and Learning Systems}, vol.~24, no.~11, pp.
  1850--1862, 2013.

\bibitem{21}
X.~Wei and C.~Li, ``The infinite student's t -mixture for robust modeling,''
  \emph{Signal Processing}, vol.~92, no.~1, pp. 224--234, 2012.

\bibitem{22}
N.~Bouguila and D.~Ziou, ``A {D}irichlet process mixture of generalized
  {D}irichlet distributions for proportional data modeling,'' \emph{IEEE
  Transactions on Neural Networks}, vol.~21, no.~1, pp. 107--122, 2010.

\bibitem{23}
X.~Wei and Z.~Yang, ``The infinite student's t -factor mixture analyzer for
  robust clustering and classification ∵,'' \emph{Pattern Recognition},
  vol.~45, no.~12, pp. 4346--4357, 2012.


\bibitem{24}
S.~P. Chatzis and G.~Tsechpenakis, ``The infinite hidden {M}arkov random field
  model.'' \emph{IEEE Transactions on Neural Networks}, vol.~21, no.~6, pp.
  1004--14, 2010.


\bibitem{Bouguila2007}
N.~Bouguila and D.~Ziou, ``High-dimensional unsupervised selection and
  estimation of a finite generalized {D}irichlet mixture model based on minimum
  message length.'' \emph{IEEE Transactions on Pattern Analysis and Machine
  Intelligence}, vol.~29, no.~10, pp. 1716--31, Aug. 2007.

\bibitem{Bouguila2012}
N.~Bouguila, ``Hybrid generative/discriminative approaches for proportional
  data modeling and classification,'' \emph{IEEE Transactions on Knowledge and
  Data Engineering}, vol.~24, no.~12, pp. 2184--2202, July 2012.

\bibitem{14}
M.~Wedel and P.~Lenk, \emph{Markov Chain Monte Carlo}.\hskip 1em plus 0.5em
  minus 0.4em\relax Boston, MA: Springer US, 2013, pp. 925--930.

\bibitem{Robert2007}
C.~P. Robert, \emph{The Bayesian Choice: From Decision-Theoretic Foundations to
  Computational Implementation}.\hskip 1em plus 0.5em minus 0.4em\relax
  Springer-Verlag New York, 2007.

\bibitem{Pereyra2016}
M.~Pereyra, P.~Schniter, E.~Chouzenoux, J.~C. Pesquet, J.~Y. Tourneret, A.~O.
  Hero, and S.~McLaughlin, ``A survey of stochastic simulation and optimization
  methods in signal processing,'' \emph{IEEE Journal of Selected Topics in
  Signal Processing}, vol.~10, no.~2, pp. 224--241, Mar. 2016.



\bibitem{15}
M.~I. Jordan, Z.~Ghahramani, T.~S. Jaakkola, and L.~K. Saul, ``An introduction
  to variational methods for graphical models,'' \emph{Machine Learning},
  vol.~37, no.~2, pp. 183--233, 1999.


\bibitem{06}
J.~Taghia and A.~Leijon, ``Variational inference for {W}atson mixture model,''
  \emph{IEEE Transactions on Pattern Analysis and Machine Intelligence},
  vol.~38, no.~9, pp. 1886--1900, 2015.



\bibitem{Paisley2015}
J.~Paisley, C.~Wang, D.~M. Blei, and M.~I. Jordan, ``Nested hierarchical
  {D}irichlet processes,'' \emph{IEEE Transactions on Pattern Analysis and
  Machine Intelligence}, vol.~37, no.~2, pp. 256--270, Feb. 2015.

\bibitem{Fan2015}
W.~Fan and N.~Bouguila, ``Topic novelty detection using infinite variational
  inverted {D}irichlet mixture models,'' in \emph{IEEE International Conference
  on Machine Learning and Applications (ICMLA)}, Dec 2015, pp. 70--75.




\bibitem{Frigyik2010}
B.~A. Frigyik, A.~Kapila, and M.~R. Gupta, ``Introduction to the {D}irichlet
  distribution and related processes,'' Department of Electrical Engineering,
  University of Washington, Tech. Rep., 2010.


\bibitem{Paisley2009}
J.~Paisley and L.~Carin, ``Hidden {M}arkov models with stick-breaking priors,''
  \emph{IEEE Transactions on Signal Processing}, vol.~57, no.~10, pp.
  3905--3917, June 2009.

\bibitem{27}
S.~Kullback and R.~A. Leibler, ``On information and sufficiency,'' \emph{Annals
  of Mathematical Statistics}, vol.~22, no.~22, pp. 79--86, 1951.

\bibitem{16}
H.~Attias, ``A variational bayesian framework for graphical models,''
  \emph{Advances in Neural Information Processing Systems}, vol.~12, pp.
  209--215, 2000.

\bibitem{30}
A.~P. Dempster, N.~M. Laird, and D.~B. Rubin, ``Maximum likelihood from
  incomplete data via {EM} algorithm,'' \emph{Journal of the Royal Statistical
  Society}, vol.~39, pp. 1--38, 1977.

\bibitem{S2}
A.~Petronis, ``Epigenetics as a unifying principle in the aetiology of complex traits and diseases'',~\emph{Nature}. No.465, pp. 721-727, 2010.

\bibitem{S3}
J.~G.~Liao, Y.~Lin, Z.~E.~Selvanayagam, and W.~J.~Shih, ``A mixture model for estimating the local false discovery rate in DNA microarray analysis'',~\emph{Bioinformatics}. Vol. 20, No. 16, pp. 694-701, 2004.

\bibitem{S4}
J.~Sandoval, H.~Heyn, S.~Moran, J.~Serra-Musach, M.A.~Pujana, M.~Bibikova, and M.~Esteller, ``Validation of a DNA methylation microarray for $450,000$ CpG sites in the human genome'', Journal of Epigenetics, 6, 692-702, 2011.


\bibitem{S5}
K.~D.~Siegmund, P.~W.~Laird, and I.~A.~Laird-Offringa, ``A comparison of cluster analysis methods using DNA methylation data'',~\emph{Bioinformatics}, Vol.20 No. 12, pp. 1896-1904, 2004.

\bibitem{S6}
P.~Du, X.~Zhang, C.~C.~Huang, N.~Jafari, W.~A.~Kibbe, L.~Hou, and S.~M. Lin, ``Comparison of Beta-value and M-value methods for quantifying methylation levels by microarray analysis'',~\emph{BMC Bioinformatics}. No. 11, 2001.


\bibitem{S7}
Z. Ma, A.E. Teschendorff, H. Yu, J. Taghia, J. Guo, ``Comparisons of Non-Gaussian Statistical Models in DNA Methylation Analysis'',~\emph{International Journal of Molecular Science}. No. 15, pp. 10835-10854 ,2014.

\bibitem{S33}
K.~Laurila, B.~Oster, C.~Andersen, P.~Lamy, T.~Orntoft, O.~Yli-Harja, and C.~Wiuf, ``A beta-mixture model for dimensionality reduction, sample classification and analysis'',~\emph{BMC Bioinformatics}, 2011.

\bibitem{S8}
Z.~Ma and A.~Leijon, ``Bayesian Estimation of Beta Mixture Models with Variational Inference'',~\emph{IEEE Transactions on Pattern Analysis and Machine Intelligence}. No. 33, pp. 2160-2173, 2011.

\bibitem{S9}
Z. Ma and A.E. Teschendorff, ``A Variational Bayes Beta Mixture Model for Feature Selection in DNA Methylation Studies'',~\emph{ Journal of Bioinformatics and Computational Biology}. Vol.11 No.4, pp.19, 2013.

\bibitem{S10}
C.~M. Bishop, ``Pattern Recognition and Machine Learning'',~\emph{Berlin/Heidelberg}, Germany, 2006.

\bibitem{S11}
J.~S. Liu, ``The Collapsed Gibbs Sampler in Bayesian Computations with Applications to a Gene Regulation Problem'',~\emph{ American Statistical Association}. Vol. 89 No. 427, pp. 958-966. 1994.


\bibitem{S13}
B.~A. Flusberg, D.~R. Webster, J.~H. Lee, K.~J. Travers, E.~C. Olivares, T.~A. Clark, J.~Korlach, S.~W. Turner, ``Direct detection of DNA methylation during single-molecule, real-time sequencing'',~\emph{Nature Methods}. No. 7, pp. 461-465, 2010.

\bibitem{S14}
G.~Wang, A.~V.~Kossenkov, M.~F. Ochs, ``LS-NMF: A modified non-negative matrix factorization algorithm utilizing uncertainty estimates'',~\emph{BMC. Bioinform}. No. 7, pp. 175, 2006.


\bibitem{b1}
M.~K.~K. Leung, H.~Y. Xiong, L.~J. Lee, and B.~J. Frey, ``Deep learning of the tissue-regulated splicing code'',~\emph{Bioinformatics}. Vol.30, pp. 121-129, 2014.

\bibitem{b2}
N.~Y. Hammerla, J.~M. Fisher, P.~Andras, L. Rochester, R. Walker, and T. Pl\"{o}tz, ``PD Disease State Assessment in Naturalistic Environments using Deep Learning'',~\emph{Twenty-Ninth AAAI Conference on Artificial Intelligence}, 2015

\bibitem{b3}
S.~Jirayucharoensak, S.~Pan-Ngum, and P.~Israsena,  ``EEG-Based Emotion Recognition Using Deep Learning Network with Principal Component Based Covariate Shift Adaptation'',~\emph{The Scientific World Journal}, 2014.

\bibitem{S15}
S.~Jurgen, ``Deep Learning in Neural Networks: An Overview'',~\emph{Technical report, IDSIA}, 2014.

\bibitem{S16}
G.~E. Hinton and R.~R. Salakhutdinov, ``Reducing the Dimensionality of Data with Neural Networks'',~\emph{Science}. Vol. 313, No. 5786, pp. 504-507, 2006.

\bibitem{S17}
D.~Ciresan, U.~Meier, and J.~Schmidhuber, ``Multi-column Deep Neural Networks for Image Classification'',~\emph{Computer Vision and Pattern Recognition}. pp. 3642-3649, 2012.


\bibitem{S18}
R.~Collobert, J.~Weston, L.~Bottou, M.~Karlen, K.~Kavukcuoglu, and P.~Kuksa, ``Natural Language Processing (Almost) from Scratch'',~\emph{Journal of Machine Learning Research}. Vol. 12, pp. 2493-2537, 2011.

\bibitem{S19}
G.~Hinton, L.~Deng, D.~Yu, A. Mohamed, N. Jaitly, A. Senior, V. Vanhoucke, P. Nguyen, T.~S. George Dahl, and B. Kingsbury, ``Deep Neural Networks for Acoustic Modeling in Speech Recognition''~\emph{IEEE Signal Processing Magazine}. Vol. 11, pp. 82-97, 2012.


\bibitem{S20}
M.~Matusugu, M.~Katsuhiko, M.~Yusuke, and K.~Yuji, ``Subject independent facial expression recognition with robust face detection using a convolutional neural network'',~\emph{Neural Networks}. Vol.16, No.5, pp. 555-559, 2013.

\bibitem{S21}
S.~Kihyuk, Z.~Guanyu, L.~Chansoo, and L.~Honglak, ``Learning and Selecting Features Jointly with Point-wise Gated Boltzmann Machines'', in proceedings of \emph{International Conference on Machine Learning}, 2013.

\bibitem{S22}
GEO. Gene Expression Omnibus. Available online: http://www.ncbi.nlm.nih.gov/geo/

\bibitem{S23}
J.~Zhuang, A.~Jones, S.~H.~Lee, E.~Ng, H.~Fiegl, M.~Zikan, D.~Cibula, A.~Sargent, H.~B.~Salvesen, and I.~J. Jacobs, ``The dynamics and prognostic potential of DNA methylation changes at stem cell gene loci in women＊s cancer'',~\emph{PLoS Genet}. Vol.8, No. 3, 2012.

\bibitem{S24}
M.~L. Vander and G. Hinton, ``Visualizing Data using t-SNE'',~\emph{Journal of Machine Learning Research}. Vol. 9, pp. 2579-2605, 2008.

\bibitem{T1}
I.~Gashi, V.~Stankovic, C.~Leita, and O.~Thonnard, ``An Experimental Study of Diversity with Off-the-shelf AntiVirus Engines'', in proceedings of~\emph{the IEEE International Symposium on Network Computing and Applications}, 2009.

\bibitem{T2}
P.~Hamel and D.~Eck, ``Learning Features from Music Audio with Deep Belief Networks'', in proceedings of~\emph{the International Society for Music Information Retrieval Conference}. pp. 339-344, 2010.
\bibitem{T3}
A.~R. Jamieson, M.~L. Giger, K. Drukker, H. Lui, Y. Yuan and N. Bhooshan, ``Exploring Nonlinear Feature Space Dimension Reduction and Data Representation in Breast CADx with Laplacian Eigenmaps and t-SNE'',~\emph{Medical Physics}. Vol. 37, No. 1, pp. 339-351, 2010.

\bibitem{T4}
I.~Wallach and R.~Liliean, ``The Protein-Small-Molecule Database, A Non-Redundant Structural Resource for the Analysis of Protein-Ligand Binding'',~\emph{Bioinformatics}. Vol. 25, No. 5, PP. 615-620, 2009.


\bibitem{S25}
T.~Kohonen, ``Self-Organized Formation of Topologically Correct Feature Maps'',~\emph{Biological Cybernetics}. Vol. 43, No. 1, pp. 59-69, 1982.

\bibitem{S26}
Z. Ma, A. E. Teschendorff, A. Leijon, Y. Qiao, H. Zhang, and J. Guo, ``Variational Bayesian Matrix Factorization for Bounded Support Data'',~\emph{ IEEE Transaction on Pattern Analysis and Machine Intelligence}.Vol. 37, No. 4, pp. 876-889, 2015.








%
%
%
%
%
%
%
%
%
%
%
%
%
%
%
%
%
%
%
%
%
%
%
%
%
%
%
%
%
%
%

\end{thebibliography}
\end{document}